\documentclass[preprint2]{aastex}

\shorttitle{Do local analogs of LBGs exist?}
\shortauthors{Scarpa, Falomo, and Lerner}

\begin{document}


\title{Do local analogs of Lyman Break Galaxies exist?}


\author{R. Scarpa }
\affil{Instituto de astrof\'isica de Canarias, Spain}
\email{riccardo.scarpa@gtc.iac.es}

\author{R. Falomo}
\affil{Padova Observatory, Italy}
\email{renato.falomo@oapd.inaf.it}

\and

\author{E. Lerner}
\affil{Lawrenceville Plasma Physics, Inc., USA}
\email{elerner@igc.org}

\begin{abstract} The optical properties of a number of 
supercompact ultraviolet luminous galaxies (UVLG), recently discovered
in the local Universe matching GALEX and Sloan Digital Sky Survey
(SDSS) data, are discussed. Detailed re-analysis of the SDSS data for
these and other similar but nearer galaxies shows that their surface
brightness radial profile in both R and u bands is in most cases well
described by an extended disk plus a central unresolved component
(possibly a bulge).  Since the SDSS pipeline used a single disk
component to derive the half light radius of these UVLGs their size
was severely underestimated.  Consequently, the average UV surface
brightness is much lower that previously quoted casting doubts on the
claim that UVLGs are the local analogs of high redshift Lyman break
galaxies.
\end{abstract}


\section{Introduction}

\begin{figure*}
\includegraphics[angle=0,scale=.525]{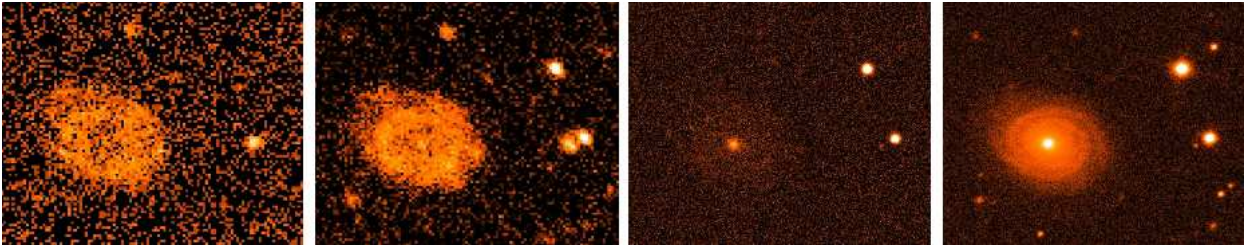}
\caption{From left to right the GALEX far-UV, GALEX near-UV, SDSS
$u$-band, and SDSS R-band images of a field galaxy located at 02:44:45
-08:09:52 This galaxy was found in the same GALEX image and not far
from object 02:45:29 -08:16:37 belonging to the Hoopes list. Note that
these are GALEX MIS images, deeper than the AIS images used by
Hoopes. The diameter of the disk of the galaxy is approximately 1
arcmin.}
\end{figure*}

Lyman Break Galaxies (LBGs; Steidel et al. 1999) are
high redshift galaxies thought to be undergoing intense star
formation. These galaxies are characterized by compact structure and
extremely high surface brightness ($>10^9$ L$_{\odot}$ kpc$^{-2}$) and
have been suggested to be the precursors of present-day elliptical
galaxies (see e.g., Giavalisco 2002). While LBGs are rather common at
z$>2.5$, it remains unclear whether local analogs of these interesting
objects do exist.

In a recent letter by Heckman et al (2005), with details given in
Hoopes et al. (2006; Paper 1 \& 2 hereafter), the finding in the local
Universe ($z<0.3$) of supercompact UV luminous ( L$_{1530}> 2 \times
10^{10}$ L$_\odot$) galaxies that share the properties of high
redshift LBGs ( having UV surface brightness above $10^9$ L$_{\odot}$
kpc$^{-2}$) was reported.  These galaxies were identified using data
from the GALEX AIS survey (Martin et al. 2005) and Sloan Digital Sky
Survey (SDSS) third data release (Abazajian et al. 2005).  In
particular, 33 galaxies were identified in paper 1 \& 2 as having UV
luminosity, size, and surface brightness adequate for being classified
as local analogs of LBGs.

In Paper 1 \& 2 the far-UV surface brightness of the 
UVLGs is not directly measured. Instead, the half-light
radii measured by SDSS in the u-band are used to estimate the surface
area of the far-UV emitting region.
 Specifically, the UV average surface brightness was derived by
dividing half the total far-UV flux (from the GALEX archive) by the
galaxy surface $\pi$r$_e^2$ , where r$_e$ is the half-light radius
from the seeing-corrected exponential model fit calculated by the SDSS
pipeline in the u-band. This procedure yielded 33 supercompact
galaxies with far-UV surface brightness I$_{1530} > 10^9$ L$_{\odot}$
kpc$^{-2}$, the alleged LBGs local analogs. For this particular set of
galaxies, the half-light radii range from $0.10 <$ r$_e$ $< 1.04$
arcsec with median value of $0.41$ arcsec. It is because of these
small radii that the surface brightness is high, the luminosities
being in no way exceptional among UVLGs.

It is worth noticing that the quoted u-band half-light radii
(Table 1 in Hoopes et al. 2006) are particularly small being in many
cases below 0.3 arcsec. This is much smaller than the typical seeing
($>0.8$ arcsec) of SDSS data (Abazajian et al. 2005) and suggests that
the SDSS automatic routine used to compute them might have failed.  To
be more precise, the SDSS pipeline is asked to fit a single disk
(exponential) model. In the case of galaxies with two clearly
distincted components (the disk and the bulge), this is inappropriate
because the algorithm will inevitably tend to fit the central part,
the most luminous one, of the surface brightness radial profile. As a
result, in a number of cases the half-light radius quoted in the SDSS
archive is possibly appropriate for the bulge while missing the faint
disk of the galaxy.

Puzzled by this thoughts and given the importance of finding local
analogs of LBGs for our understanding of the process of galaxy
formation, we were prompted to re-analyzed the SDSS images to check
whether the half-light radius in the SDSS archive provides a reliable
measure of the far-UV radii. We shall show that there are good reasons
to believe that this is not the case and that the local analogs of
LBGs, if they exist at all, have still to be found.

\begin{deluxetable}{cccccccccccc}
\tabletypesize{\scriptsize}
\tablecaption{Hoopes sample: fit to the u- and R-band data}
\tablewidth{470pt}
\tablehead{
\colhead{\#} & \colhead{RA} & \colhead{DEC} & \colhead{z} & \colhead{SDSS r$_e$} &
\colhead{log(UV$_{SB}$)} &\colhead{Best Fit} & \colhead{Core} & \colhead{Disk} &
\colhead{Disk r$_e$} & \colhead{Corrected}  & \colhead{notes}\\
\colhead{} & \colhead{} & \colhead{} & \colhead{} & \colhead{$u$-band} &
\colhead{L$_\odot$ kpc$^{-2}$} &\colhead{} & \colhead{mag} & \colhead{Total} &
\colhead{R-band} & \colhead{log(UV$_{SB}$)}  & \colhead{}\\
\colhead{} & \colhead{} & \colhead{} & \colhead{} & \colhead{arcsec} &
\colhead{} &\colhead{} & \colhead{} & \colhead{mag} &
\colhead{arcsec} & \colhead{L$_\odot$ kpc$^{-2}$}  & \colhead{}
}
\startdata
01 &  01:50:28 &  13:08:58 &  0.147 &  0.65 &  9.38 &   PSF+disk &  19.2 & 18.1  & 1.2 &  8.84&    \\
02 &  02:13:48 &  12:59:51 &  0.219 &  0.24 &  9.92 &   PSF+disk &  18.6 & 18.4  & 2.8 &  7.78&    \\
   &           &           &        &       &       &            &  19.2 & 21.5  & 2.8 &      &    \\
03 &  01:02:26 &  14:54:38 &  0.086 &  1.04 &  9.08 &     disk   &   *   & 17.5  & 1.05&  9.07&  1 \\
   &           &           &        &       &       &            &   *   & 19.4  & 0.9 &      &    \\
04 &  00:40:54 &  15:34:09 &  0.283 &  0.40 &  9.16 &     PSF    &  19.8 &   *   &  *  &  8.36&  2 \\
05 &  00:44:47 &  15:29:11 &  0.227 &  0.63 &  9.04 &   PSF+disk &  19.9 & 18.8  & 1.1 &  8.55&  3 \\
   &           &           &        &       &       &            &  20.1 & 20.4  & 1.1 &      &    \\ 
06 &  01:51:25 &  13:25:10 &  0.243 &  0.40 &  9.35 &   PSF+disk &  20.0 & 19.3  & 0.8 &  8.74&    \\
07 &  02:45:29 &$-$08:16:37&  0.195 &  0.51 &  9.17 &   PSF+disk &  19.3 & 19.2  & 0.8 &  8.77&  3 \\
08 &  08:17:22 &  46:44:59 &  0.280 &  0.76 &  9.01 &   PSF+disk &  19.2 & 20.0  & 1.3 &  8.54&  4 \\
09 &  10:51:45 &  66:06:21 &  0.170 &  0.30 &  9.64 &   PSF+disk &  19.2 & 19.4  & 1.2 &  8.43&    \\
10 &  13:53:55 &  66:48:00 &  0.198 &  0.58 &  9.31 &   PSF+disk &  19.0 & 18.8  & 0.8 &  9.03&  3 \\
11 &  09:48:24 &  61:39:56 &  0.173 &  0.35 &  9.60 &   PSF+disk &  19.1 & 20.3  & 0.9 &  8.78&  3 \\
12 &  10:12:11 &  63:25:03 &  0.246 &  0.32 &  9.35 &   PSF+disk &  19.2 & 20.5  & 2.2 &  7.67&  5 \\
13 &  08:15:23 &  50:04:14 &  0.164 &  0.66 &  9.16 &   PSF+disk &  18.7 & 17.6  & 1.0 &  8.79&    \\
14 &  04:02:08 &$-$05:06:42&  0.139 &  0.49 &  9.49 &   PSF+disk &  19.0 & 18.8  & 1.2 &  8.71&    \\
   &           &           &        &       &       &            &  19.9 & 20.6  & 1.2 &      &    \\ 
15 &  11:39:47 &  63:09:11 &  0.245 &  0.51 &  9.02 &   PSF+disk &  19.2 & 19.7  & 0.9 &  8.52&    \\
16 &  09:23:36 &  54:48:39 &  0.222 &  0.32 &  9.65 &   PSF+disk &  19.6 & 20.6  & 1.0 &  8.66&    \\
17 &  08:08:44 &  39:48:52 &  0.091 &  0.29 & 10.20 &   PSF+disk &  17.8 & 17.9  & 2.2 &  8.44&  6 \\
   &           &           &        &       &       &            &  18.2 & 20.0  & 2.1 &      &    \\
18 &  12:39:31 &  64:41:05 &  0.133 &  0.63 &  9.36 &   PSF+disk &  18.8 & 18.8  & 1.1 &  8.87&  7 \\
19 &  11:33:03 &  65:13:41 &  0.241 &  0.10 & 10.80 &   PSF+disk &  20.1 & 20.8  & 1.0 &  8.8 &    \\
20 &  20:50:00 &  00:31:24 &  0.164 &  0.21 & 10.00 &   PSF+disk &  18.8 & 19.7  & 1.1 &  8.56&  3 \\
21 &  23:25:39 &  00:45:07 &  0.277 &  0.25 &  9.67 &   PSF+disk &  19.9 & 21.4  & 1.2 &  8.30&    \\
22 &  21:45:00 &  01:11:57 &  0.204 &  0.23 &  9.82 &   PSF+disk &  18.9 & 19.5  & 1.0 &  8.54&    \\
23 &  23:07:03 &  01:13:11 &  0.126 &  0.31 &  9.93 &   PSF+disk &  18.5 & 19.8  & 0.9 &  9.00&  8 \\
24 &  03:28:45 &  01:11:50 &  0.142 &  0.70 &  9.11 &   PSF+disk &  19.6 & 18.6  & 0.9 &  8.89&    \\
25 &  09:21:59 &  45:09:12 &  0.235 &  0.41 &  9.65 &   PSF+disk &  18.6 & 18.6  & 1.6 &  8.46&  9 \\
26 &  09:51:37 &  48:39:41 &  0.135 &  0.79 &  9.04 &   PSF+disk &  19.7 & 19.3  & 0.7 &  9.14&    \\
27 &  08:21:37 &  37:10:46 &  0.284 &  0.43 &  9.47 &   PSF+disk &  19.6 & 19.3  & 0.5 &  9.33&    \\
28 &  09:26:00 &  44:27:36 &  0.181 &  0.34 &  9.91 &   PSF+disk &  19.2 & 19.5  & 0.7 &  9.28&    \\
29 &  09:54:34 &  51:35:08 &  0.130 &  0.67 &  9.60 &   PSF+disk &  18.5 & 18.2  & 0.9 &  9.34&    \\
30 &  23:18:12 &$-$00:41:26&  0.252 &  0.61 &  9.29 &   PSF+disk &  19.3 & 18.7  & 1.2 &  8.70&    \\
   &           &           &        &       &       &            &  20.3 & 20.5  & 1.2 &      &    \\
31 &  00:10:09 &$-$00:46:03&  0.243 &  0.36 &  9.37 &   PSF+disk &  20.2 & 19.6  & 1.0 &  8.48& 10 \\ 
   &           &           &        &       &       &            &  21.1 & 21.6  & 1.1 &      &    \\
32 &  00:55:27 &$-$00:21:48&  0.167 &  0.28 &  9.89 &   PSF+disk &  18.4 & 19.9  & 0.9 &  8.87&    \\
33 &  23:53:47 &  00:54:02 &  0.223 &  0.60 &  9.01 &   PSF+disk &  19.9 & 20.2  & 0.7 &  8.87&    \\
\enddata
\tablecomments{\\
The first line for each source refers to the R-band data, while when present the 
second line gives the best fit in the u-band.
Uncertainties on best fit parameters in the R-band are $\pm 0.1$ magnitudes and $\pm 0.1$ arcsec, or smaller.
In the u-band errors are $\pm 0.1$ magnitudes for the core and $\pm 0.2$ mag and $\pm 0.2$ arcsec
for the disk.\\
1: Problem in GALEX image. Source right on the edge of the field of view.\\
2: Unresolved. To compute the surface brightness the radius was set to 1 arcsec, similar to 
the atmospheric seeing and equal to the median of the sample.\\ 
3: Double.\\
4: Bad Fit. Core with extremely boxy isophotes and an off-center light pick.\\
5: Close to bright star.\\
6: Triple. Also acceptable fit with a de Vaucouleurs law with r=1.3 arcsec.\\
7: Object  right on-top of the diffraction figure of a nearby
  bright star.  The shape of the low intensity isophotes is strongly
  affected by this diffuse light in the north-east side of the galaxy.\\
8: Elongated, possibly double.\\
9: Head-tail morphology. Possibly embedded companion at 1.2 arcsec.\\
10 Head-tail morphology. Possibly embedded companion at 0.9 arcsec.\\
}
\end{deluxetable}

\begin{figure*}
\includegraphics[angle=0,scale=.95]{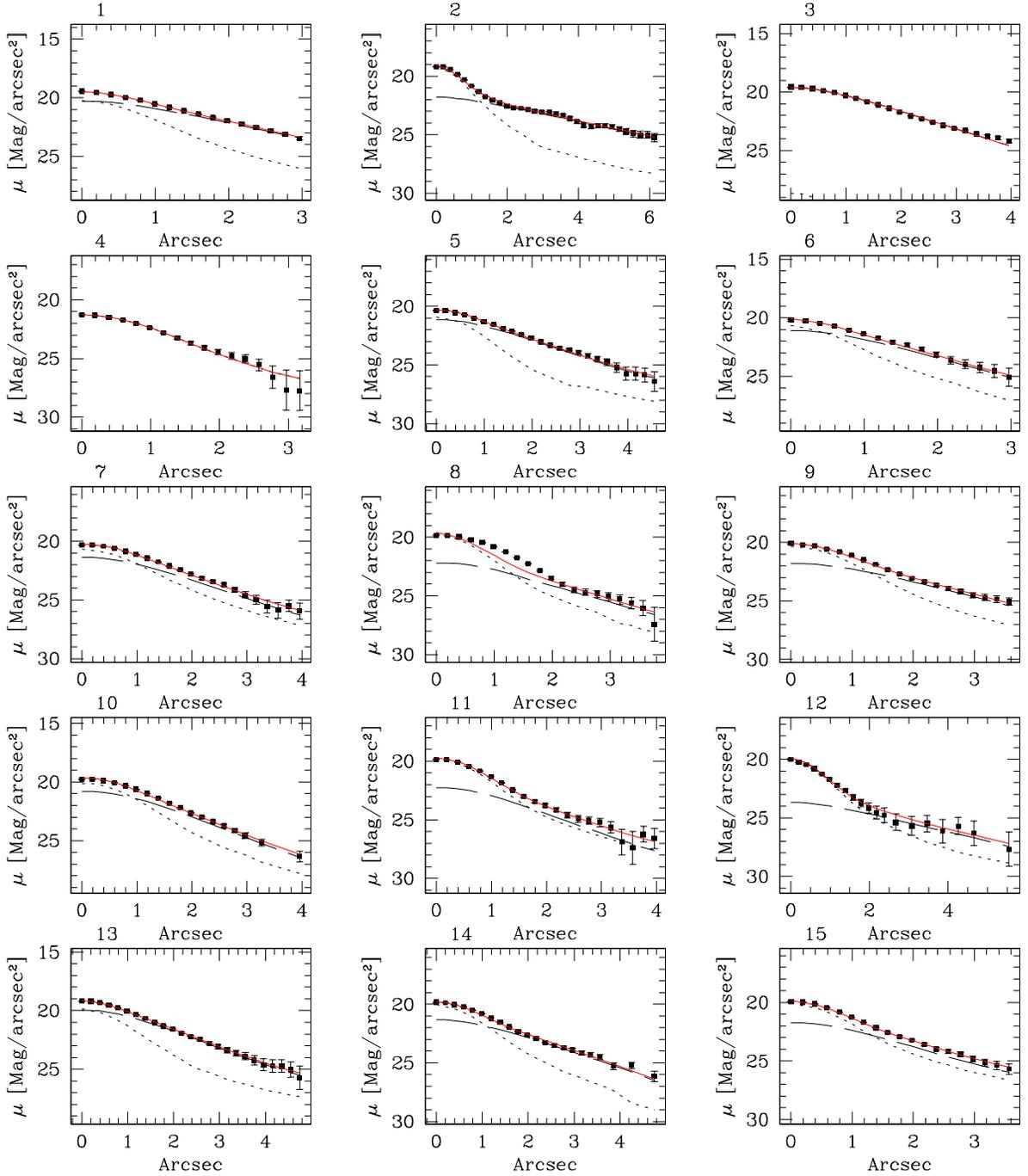}
\caption{Best fit of the average R-band surface brightness radial profile (points
with error bars) for all object in Table 1. The observed
radial profile is shown by the points with errorbars. The PSF model is
shown by a dotted line, while the disk model convolved with the PSF is
shown by a dashed line.  The sum of the two components is shown by the
solid line.}
\end{figure*}

\addtocounter{figure}{-1}
\begin{figure*}
\includegraphics[angle=0,scale=.95]{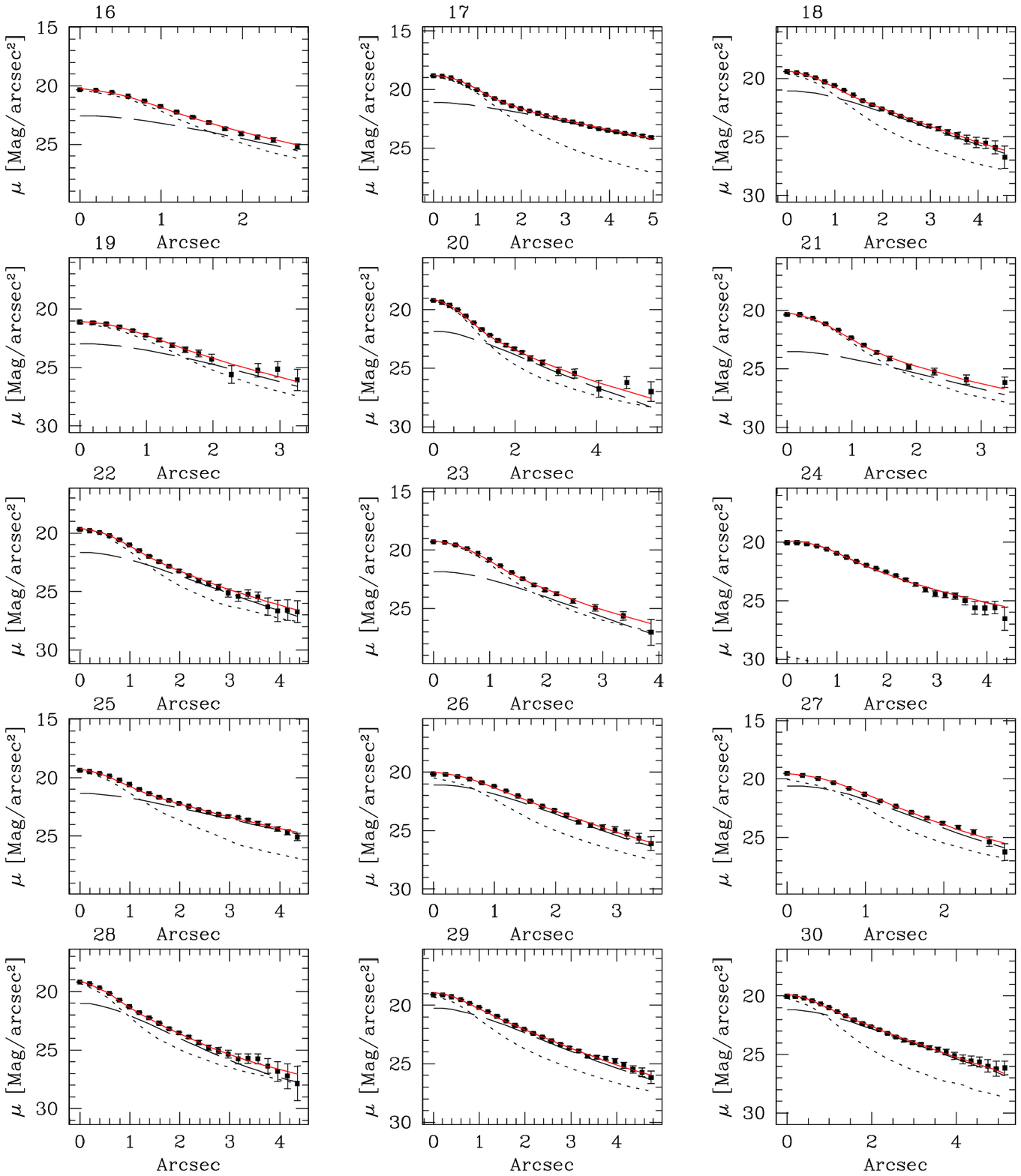}\\
\caption{Continued}
\end{figure*}

\addtocounter{figure}{-1}
\begin{figure*}
\includegraphics[angle=0,scale=.95]{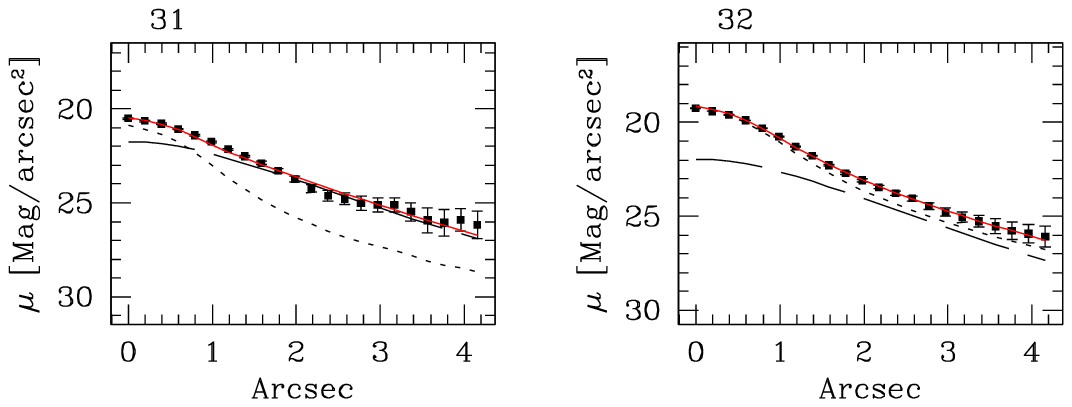}\\
\caption{Continued}
\end{figure*}

\section{The Hoopes et al. analysis revisited}

To identify in the local Universe the analogs of LBGs both 
flux and size in the far-UV of candidates are needed. Until recently
neither of the two were available for large samples of data. GALEX has
provided far-UV (1530 \AA) fluxes for thousands of galaxies, allowing
the identification of 215 galaxies as luminous as LBGs (Paper 1 \&
2). With luminosity at 1530 \AA\ L$_{1530}>2\times 10^{10}$
L$_{\odot}$, these galaxies represent the high-luminosity end of the
UV luminosity function of non-active galaxies.

While many of the galaxies detected by GALEX are resolved and a direct
measure of their UV size is possible, the apparent size of these 215
UV-luminous galaxies is in most cases too small to be resolved.  Thus,
to estimate their surface brightness the size derived at a different
wavelength was used in Paper 1 \& 2, under {\it the assumption that
the light distribution does not change significantly between the two
bands}. This is a risky assumption, for the UV and optical light
distribution in normal galaxies is markedly different. This is nicely
illustrated, for instance, in the case of the Andromeda galaxy
(Thilker et al. 2005).  Another example that allow a quantitative
discussion of the SDSS pipeline products is shown in Fig. 1, where we
compare the GALEX and SDSS images of a large field galaxy. While in
the optical band the emission is dominated by the central bulge, in
the UV there is almost no light coming from the center. Moreover, in
the shallower u-band image (in the Sloan survey the exposure time is
the same for both the u and R bands), the disk of the galaxy is mostly
undetected. In these two images the SDSS pipeline fits two completely
different light distributions, yielding best fit exponential models
with r$_e$ = 2.4 and 12.9 arcsec in the u and R bands,
respectively. Applying the procedure followed in Paper 1 \& 2 the UV
surface brightness is overestimated by almost 30 times. Clearly, in
the u-band the pipeline has fitted only the central core of the galaxy
while the UV emission comes from the extended disk. We learn here that
(a) the deeper R-band image should be preferred to characterize the
optical size of the galaxy, (b) the UV and optical size of the $disk$
is basically the same, and (c) if a prominent bulge exists a single
component fit is inappropriate even to describe the optical data, let
alone the far-UV.

Thus, we retrieved from the SDSS archive the raw images for all the 33
objects reported as having surface brightness above $10^9$ L$_{\odot}$
kpc$^{-2}$ in the Hoopes et al. list, and re-analyze the average
surface brightness profile extracted from both the u- and R-band
images allowing for the presence of multi-components.  The fitting
procedure was as follow. First, for each object a model of the PSF was
created averaging the radial profile of a number of bright stars in
the same frame of the object. Then this model was used to convolve the
galaxy profile, either an exponential (disk) or a de Vaucouleurs
law. The best fit was then determined by chi-squared minimization.
With only two exceptions, it was found that the surface brightness
radial profile extracted from the R-band data is well described as the
sum of a central unresolved point source, possibly the bulge of the
galaxy, plus an extended disk. Results are given in Table 1 and Figure
2. In no case could a de Vaucouleurs law give better results than a
disk model.  Object \# 3 is the only one for which we can reproduce
the SDSS fit with a single disk and no central component. In one case
(object \# 4) because of the poor atmospheric seeing ($1.5$ arcsec)
the source is found fully consistent with the PSF profile and therefore
the half-light radius quoted in the SDSS archive is meaningless.

The u-band data are considerably shallower than the
R-band data. Because of this only in seven cases the disk is large and
bright enough to be detected also in the u-band.  Results of the best
fit are reported in Table 1 (second entry for each object) and shown
in Fig. 3.  The surface brightness profile of the remaining 26 sources
is dominated by the unresolved core while the disk is virtually
undetected. As a consequence, objects are either unresolved or
marginally resolved and the half-light radius derived by the SDSS
pipeline is highly uncertain as already pointed out in Paper 1 \& 2.
As a whole we see that in all cases where a comparison is possible,
the best fit models from the two bands are fully consistent with each
other directly showing that the light distribution in the two bands is
very similar and either of the two can be used to characterize the
structure of the object.

\begin{figure*}
\includegraphics[angle=0,scale=.95]{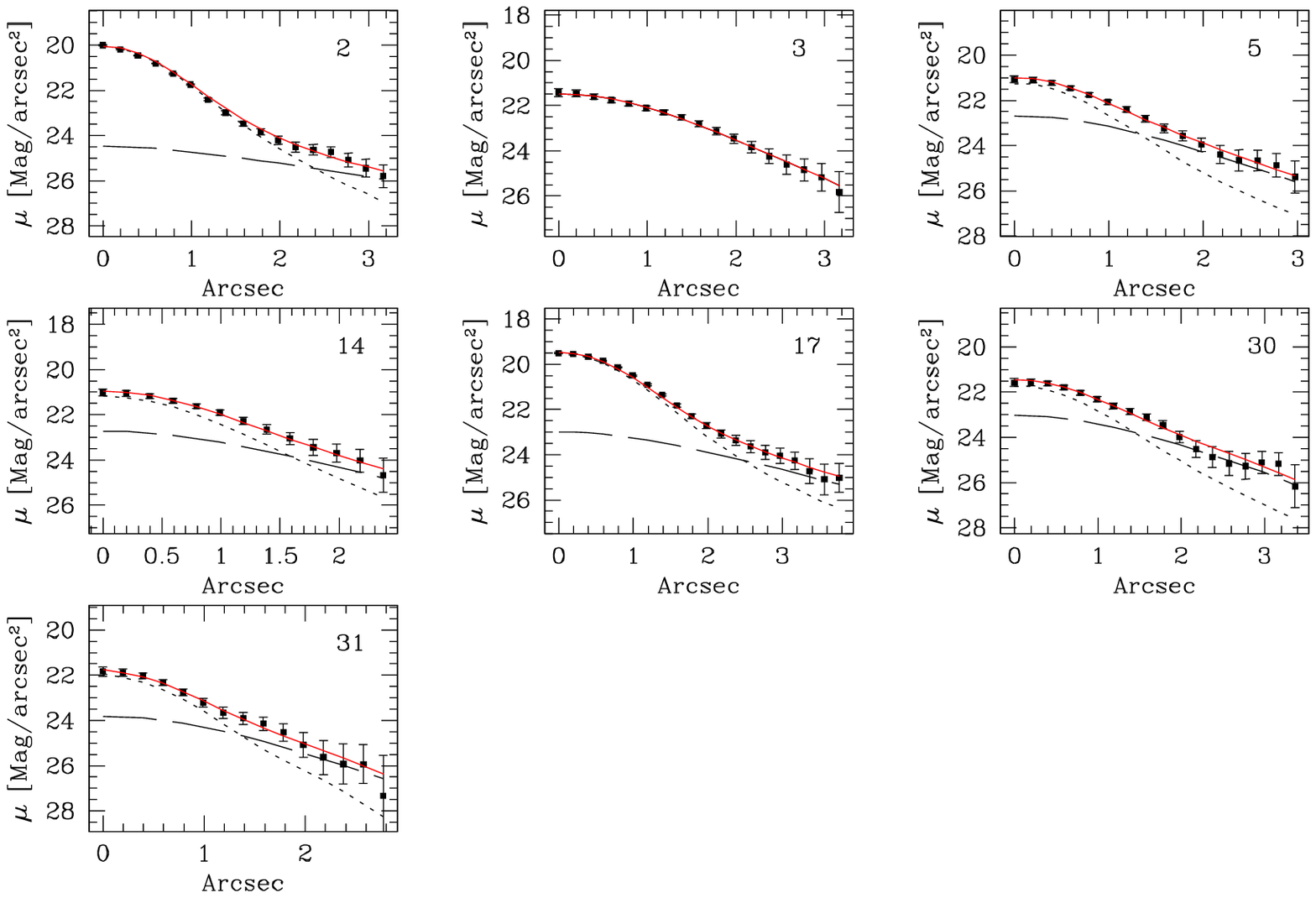}
\caption{Best fit of the average u-band surface brightness radial
profile for the brighter disks. Symbols as in figure 4.  Best fit
parameter are in Table 1.}
\end{figure*}

\begin{figure}
\includegraphics[angle=0,scale=.4]{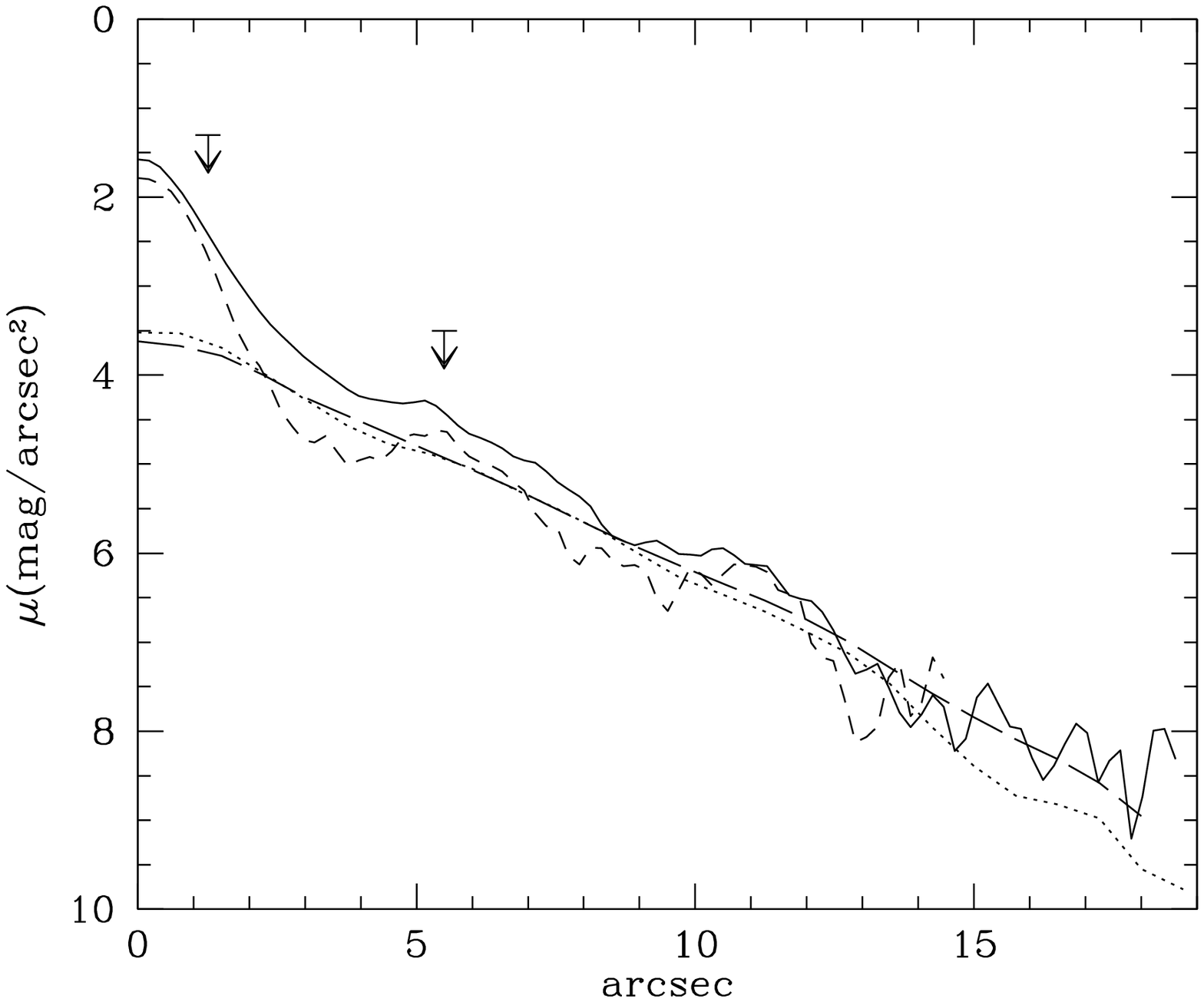}
\caption{ Comparison of the average surface brightness radial profiles
from varius bands for the galaxy at 232207.9$-$000314 (last entry in
Tab. 2). Profiles were stacked one on the other (the flux scale is
arbitrary) to evidentiate the similarities at large radii.  Different
lines show the R-band (solid), the u-band (dashed), near-UV (dotted),
and far-UV (long dashed) profiles.  The emission from the bulge
component is dominant in u- and R-band up to $\sim 3$ arcsec, while it
is negligible in the UV. The disk is detected and has the same scale
length of 5.5 arcsec (6 kpc) in all four bands. The two arrows
indicates the u-band and far-UV half-light radii.}
\end{figure}

\section{Discussion}

We have shown that the surface brightness profile of these 33
supercompact UVLGs cannot be described by a single disk model.
This result has important implication for their
identification as LBGs analogs in the local Universe.

When studying high-z LBGs the half-light radius -- not the scale
length of the disk -- is used to compute their far-UV average surface
brightness. Due to the high redshift of LBGs, the rest frame far-UV
emission is directly accessible in the optical band with adequate
resolution and thus buth luminosity and size are derived at the same
wavelength.

The use of the half-light radius then provides a model independent
measure of the average surface brightness.  On the other hand, in the
case of UVLGs the situation is different because due to the low
spatial resolution of GALEX one has to use luminosities and radii from
two different bands and the use of the half-light radius might not be
adequate.  The problem is serious indeed because our re-analysis of
the R-band and u-band data of all the 33 alleged supercompact UVLGs
has shown that most of these galaxies cannot be described with a
single disk model.  Instead, we found a significant fraction of their
total flux is due to an unresolved core, most probably the bulge. In
normal nearby galaxies stars belonging to the bulge do not contribute
significantly to the far-UV emission.  It is from the disk that most
of the far-UV radiation comes from (see discussion above).  Thus to
describe the far-UV light distribution starting from data taken in the
optical band it seems appropriate to use the scale length of the
extended disk -- not the half-light radius of the whole object -- in
order to isolate and remove the bulge component.

Now the question remains of whether this result, valid for normal
galaxies, is also applicable to supercompact UVLGs that are peculiar in
a number of ways.  To investigate this point we searched the GALEX
medium imaging survey (MIS) archive to find nearby, non-interacting,
resolved (having stellarity\footnote{The stellarity index is produced
by the Sextractor program and differentiates galaxies from unresolved
objects} $<$0.45 and half light radius $>$4.5 arcsec) SDSS galaxies for
which a direct measurement of the far-UV surface brightness can be
obtained.

\begin{figure}
\includegraphics[angle=0,scale=.6]{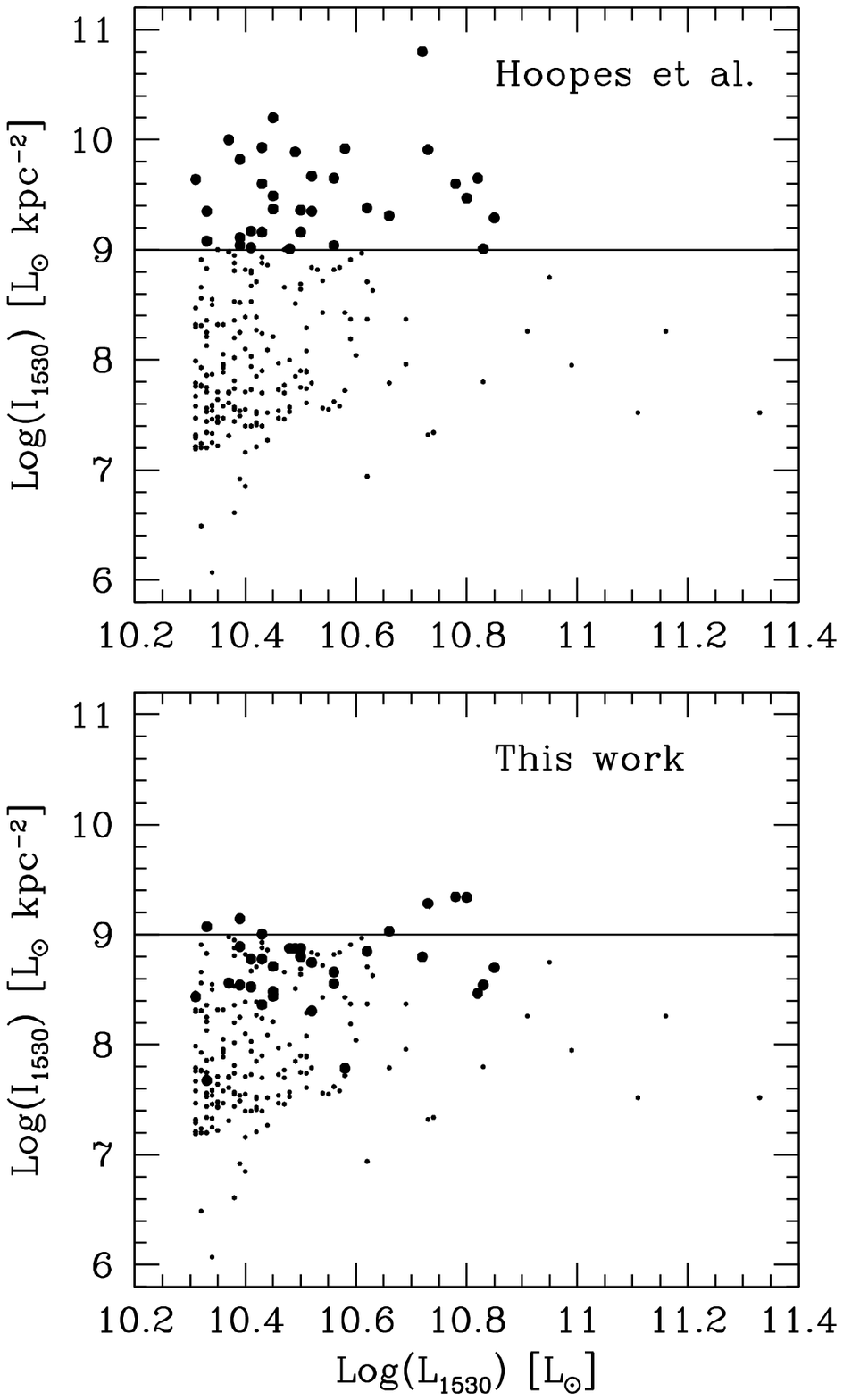}
\caption{{\bf Upper panel:} Distribution in the luminosity -- surface
brightness plane of the 215 UV luminous galaxies reported by Hoopes et
al.  According to Hoopes et al.  the region of LBGs, above
$log(I_{1530})>10^9$ L$_\odot$ kpc$^{-2}$, is populated by 33 objects
(big dots).  {\bf Lower panel:} Same as above but using our revised
surface brightness of the alleged 33 LBGs analog (big dots). In this
case the LBGs region is almost empty.}
\end{figure}

We further selected objects with color (u-R)$<$1.0, as is the case for
most objects in the Hoopes sample, and far-UV luminosity above
$10^{10}$ L$_\odot$.  This resulted in only 4 galaxies (Table 2) with
color and luminosity comparable to the one of supercompact UVLGs but
close enough to be resolved by GALEX.  In all four cases the u-band
half-light radius as derived by the SDSS pipeline is significantly
smaller than that in the far-UV. In particular, two of them are quite
compact in the u-band and their far-UV surface brightness (as
estimated by the Hoopes method) is brighter than $10^9$
L$_\odot$/kpc$^2$. Therefore they are borderline examples of the
supercompact UVLGs.  

In reality, the direct measurement of the far-UV
half-light radius show that both galaxies are quite large
and their far-UV surface brightness is below $8\times10^7$
L$_\odot$ kpc$^{-2}$.  The reason for this systematic difference
between far-UV and u-band size is the different distribution of the
stars emitting in the two bands.  In the u-band stars from both the
disk and the bulge contribute to the observed luminosity, while
virtually only the stars in the disk contribute to the far-UV emission
(Fig. 4).  Thus when galaxies like those of the Hoopes sample, but
merely closer, are observed, the u-band half light radius is typically
far smaller than the far-UV radius.

We conclude that the optical scale length of the extended disk -- not
the half-light radius used in paper 1 \& 2 -- provides the best,
thought imperfect, description of the far-UV light distribution.  This
implies that the size of the far-UV emitting region is much larger
than the value used by Hoopes, with radii ranging from 0.5 to 2.8
arcsec, with median of 1.0 arcsec.  On average this is a factor 2.5
larger than the values used by Hoopes et al., corresponding to a
reduction of 6.25 times of the surface brightness.  The most dramatic
changes, up to two orders of magnitude, occur for the objects with the
smaller radii as according to our analysis all objects with radius
smaller than 0.4 arcsec are gone.  As a whole our estimate of the
far-UV surface brightness (Tab. 1) is below the threshold for
classifying these objects as supercompact UV luminous galaxies
(Fig. 5).  The present data therefore do not support the claim of the
existence in the local universe of galaxies with properties similar to
the one of high redshift LBGs.

\begin{deluxetable}{crccccccccc}
\tabletypesize{\scriptsize}
\tablecaption{The blue sample of galaxies}
\tablewidth{400pt}
\tablehead{
\colhead{RA}& \colhead{DEC}& 
                 \colhead{z}&
                       \colhead{m$_{_{FUV}}$}&   
                                \colhead{m$_u$}&  
                                         \colhead{r$_{_{FUV}}$}& 
                                                 \colhead{r$_u$}& 
                                                         \colhead{(u$-$R)}&  
                                                             \colhead{log L$_{_{FUV}}$}&      
                                                                     \colhead{log$\Sigma_u$}&       
                                                                        \colhead{log$\Sigma_{_{FUV}}$}
}
\startdata
085625.2& $  513148 $& 0.083 &  19.3 &  18.6  &   4.64 &   3.63  &   0.98 &  10.1 &  7.75 &  7.54\\
090031.0& $  552208 $& 0.076 &  19.2 &  19.8  &   4.69 &   0.84  &   0.88 &  10.0 &  9.06 &  7.57\\
143753.8& $  050046 $& 0.087 &  19.2 &  18.1  &   5.31 &   3.00  &   0.85 &  10.1 &  7.93 &  7.43\\
232207.9& $ -000314 $& 0.057 &  17.9 &  17.8  &   6.02 &   1.26  &   0.92 &  10.3 &  9.20 &  7.84\\
\enddata
\tablecomments{
Columns give: Right ascension and Declination for the 2000 equinox; (3)
redshift; (4) far-UV magnitude as from GALEX archive; (5) Petrosian
u-band magnitude from SDSS archive; (6) far-UV half light radius from GALEX archive in
arcsec; (7) u-band SDSS effective radius from the seeing-corrected exponential model fit 
in arcsec; (8) (u-R) color in magnitudes; (9) far-UV
luminosity in solar units; (10) Average surface brightness derived
from r$_u$, in L$_\odot$ kpc$^{-2}$; (11) true far-UV surface brightness derived from r$_{FUV}$, in
L$_\odot$ kpc$^{-2}$.}
\end{deluxetable}


{\it Facilities:} \facility{GALEX}, \facility{SDSS}, 
\facility{HST (MUST archive)}.

\end{document}